\documentclass[aps,longbibliography,showpacs,twocolumn,superscriptaddress]{revtex4-1}
\usepackage{amsmath,amssymb,amsfonts,bm}
\usepackage{graphicx}
\usepackage{epstopdf}
\usepackage{dcolumn}
\usepackage{mathrsfs}
\usepackage{bbold}
\usepackage{dsfont}
\usepackage{float}
\usepackage[colorlinks=true,linkcolor=blue,citecolor=blue, urlcolor=blue,bookmarks=false]{hyperref}
\usepackage{tgtermes}
\usepackage{multirow}
\usepackage{bbm}
\usepackage{wasysym}
\usepackage{newtxtext}
\usepackage[varvw]{newtxmath}
\usepackage{hyperref}
\usepackage{array}
\usepackage{bm}
\usepackage{color}
\usepackage{float}
\usepackage{graphicx}
\usepackage{natbib}
\usepackage{newtxtext}
\usepackage{newtxmath}
\usepackage{mathrsfs}
\usepackage{amsmath,amssymb}
\usepackage{verbatim}
\usepackage{amsfonts}
\usepackage{dcolumn}
\usepackage{ulem}
\usepackage{physics}
\usepackage{subfigure}

\newcommand{\beq}{\begin{eqnarray} }
\newcommand{\eeq}{\end{eqnarray} }
\newcommand{\Beq}{\begin{eqnarray*} }
\newcommand{\Eeq}{\end{eqnarray*} }
\newcommand{\Bmat}{\left(\begin{matrix}}
\newcommand{\Emat}{\end{matrix}\right)}

\begin{document}

\title{Proposal of realizing the Sachdev-Ye-Kitaev model from a Kitaev Spin Chain}

\author{Hanyuan Zuo}
\affiliation{School of Physics and Beijing Key Laboratory of Opto-electronic Functional Materials and Micro-nano Devices, Renmin University of China, Beijing, 100872, China}
\affiliation{Key Laboratory of Quantum State Construction and Manipulation (Ministry of Education), Renmin University of China, Beijing, 100872, China}
\author{Zheng-Xin Liu}
\email{liuzxphys@ruc.edu.cn}
\affiliation{School of Physics and Beijing Key Laboratory of Opto-electronic Functional Materials and Micro-nano Devices, Renmin University of China, Beijing, 100872, China}
\affiliation{Key Laboratory of Quantum State Construction and Manipulation (Ministry of Education), Renmin University of China, Beijing, 100872, China}

\date{\today}

\begin{abstract}
The Sachdev-Ye-Kitaev (SYK) model is zero-dimensional model simulating quantum chaos using interacting Majorana fermions.Previously proposals have been made to realize the SYK model in fermionic systems that can support majorana zero modes. In this work, we simulate the SYK model in a Kitaev spin chain, where zero modes also exist in the ground state from the fermionic representation of spins.  By adding other interactions as perturbations, we lift the degenerate majorana zero modes and obatin an effective SYK interactions. Although the majorana fermions are nonlocal in forms of spin operators, we show that the out-of-time-ordered correlator (OTOC) of the majorana fermions can be approximately simulated in forms of spin operators. Our proposal sheds light in simulating the fermonic SKY model and the related physics in purely bosonic systems.

\end{abstract}

\maketitle

\section{Introduction}

The Sachdev-Ye-Kitaev (SYK) model\cite{Kitaev2015} is proposed to study quantum chaos in zero dimension. It contains $N$ Majorana fermions with random $q$-fermion interactions with $q$ an even number. It is a simple variant of the model introduced by Sachdev and Ye\cite{Sachdev1993}. The SYK model is nearly conformally invariant and exactly solvable in the large $N$ limit. Furthermore, the SYK model is shown to exhibit an interesting relationship with the horizon of an extremal black hole in two-dimensional anti-de Sitter (AdS2) space\cite{SS2010}. This connection provides a rare example of holographic duality between a solvable quantum mechanical model and gravitational dynamics. 

Out-of-time-order correlators (OTOCs), the physics of a perturbation operator observed by another operator after some time, 
which was used to describe quantum chaos, have been used to study the SYK model. OTOCs were regarded as important observables in the context of AdS/CFT correspondence or quantum gravity\cite{MS2016}. One advantage of this quantity is that, according to AdS/CFT duality, its holographic image is extremely clear, i.e., the scattering behavior of shock waves near the black hole horizon. This was already well-studied in the 1980s, which led to a series of studies on holographic quantum chaos initiated by Douglas Stanford, Steven Shenker, and others\cite{SD2014}.
Calculating OTOCs in the context of quantum information around the black hole horizon results in an exponential divergence, leading to the proposal of a quantum Lyapunov exponent\cite{MJSD2016}, analogous to classical chaos, to measure quantum chaos. This Lyapunov exponent has an upper bound, which black holes precisely achieve, making black holes the most chaotic objects in the universe. Interestingly, the SYK model in condensed matter also precisely matches this upper bound. Therefore, the saturation of the quantum Lyapunov bound indicates that the SYK model describes quantum black holes through AdS/CFT correspondence. 

Since its proposal, plenty of research interest had been sparked in studying the SYK model and its applications in and quantum gravity, high-energy physics, and condensed matter physics\cite{SS2015}. Especially, the realization of the SYM model will make it possible to simulate quantum chose and black holes in laboratoies. In 2017, it was proposed that the SYK model can be simulated in superconductors\cite{Pikulin2017}. However, since the superconductor is essentially a fermionic system, two-body coupling terms was not suppressed in the effective interactions, indicating that the effective model is not truly a SYK. 


\begin{figure}[H]
   \centering
    \includegraphics[scale = 0.28]{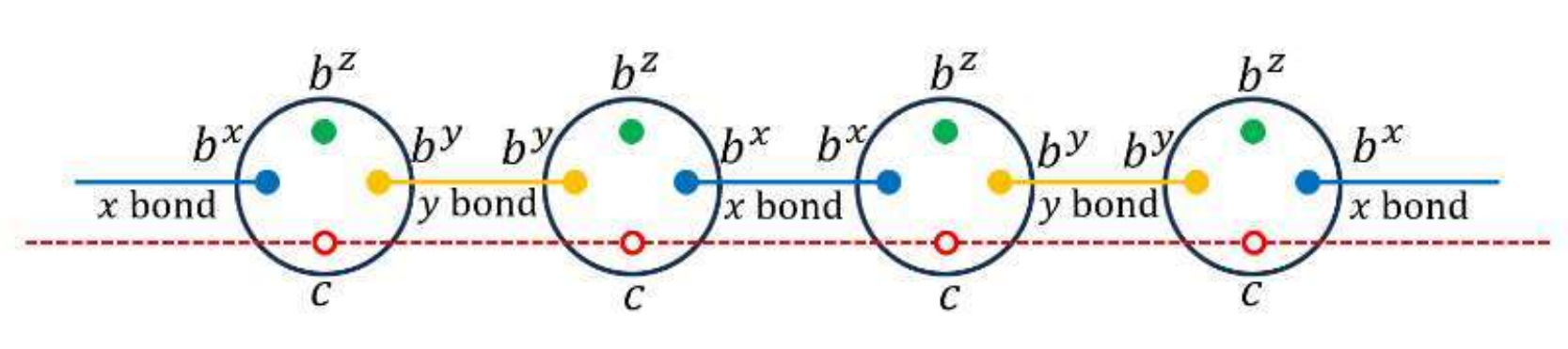}
    \caption{One-Dimensional Kitaev Chain}
    \label{fig1}
\end{figure}

In the present work, we propose an investigating scenario to realize the SYK model genuinely in a bosonic system. It is based on the Kitaev model 
which supports quantum spin liquid ground state.\cite{Lee2020} The Kitaev model on the Honeycomb lattice contains strongly anisotropic and bond-dependent interactions. In the fermionic representation of the spins, the model is exactly solvable containing localized fermions and  interacting majorana fermions.\cite{Wang20202} In the limit of extremely weak interactions on certain bonds, the honeycomb lattice Kitaev model reduces to a one-dimensional Kitaev chain (see Fig.\ref{fig1})\cite{Wang2019}\cite{Wang2020} which is still exactly solvable and contains local majorana zero modes in the ground state.  \cite{Hwang2022}Then by introducing an additional spin-interactions as perturbations, the exponentially degenerate ground state will be lifted and an pure effective SYK interactions between the local majorana modes are obtained. Our proposal indicates that the fermonic SKY model and the related interesting physics, including quantum chaos and black holes, can be simulated in purely bosonic strongly correlated systems.

The rest part of the work is organized as follows. In section II, we discuss the Kitaev spin chain, its exact solution and the effect of $\Gamma$ interactions as perturbations. we illustrate that when $K>0$ the first order perturbation vanishes and we get the the form of the second order perturbation. In section III, we demonstrate that in certain parameter region the second order perturbation gives rise to an effective SYK model. It is also shown that the OTOC of the majorana fermions can be approximately simulated via the correlators of local spin operators. Section IV devotes to the conclusion and discussion.

\section{Kitaev spin Chain and purterbations}

The Kitaev-Gamma chain for spin-1/2 spins is written as follows:
\begin{eqnarray}
    H=\sum_{\langle i j\rangle \in \gamma \rm{-bond }}\big[K S_{i}^{\gamma} S_{j}^{\gamma}+\Gamma\big(S_{i}^{\alpha} S_{j}^{\beta}+S_{i}^{\beta} S_{j}^{\alpha}\big)\big]
\end{eqnarray}

Here, i and j denote two nearest-neighbor lattice sites, $\gamma=x$(or $y$) corresponds to the spin direction related to the $\gamma$ bond, and$\alpha,\beta=y,z$(or $z,x$) are the other two spin directions different from $\gamma$ on the $\gamma$ bond. In this model, $K= \cos\phi, \Gamma = \sin\phi$. We study the region close to the K point, i.e., $\phi \to 0$, so that $\Gamma$ can be considered a small quantity, and the $\Gamma$ interaction term can be treated as a perturbation.The K-Gamma model has been widely studied in 1- and 2-dimensions \cite{Rau2014}\cite{Gohlke2018}\cite{Gordon2019}


\subsection{The Kitaev spin chain}


We start by analyzing some useful conclusions from the Kitaev spin chain. Consider a one-dimensional Kitaev chain with N lattice sites, where each site has a spin-1/2. We represent the spin operators using Majorana operators $b^x,b^y,b^z,c$, where the constraint $b^x b^y b^z c=1$ can be seen as a $\mathbb{Z}_2$ gauge transformation. We combine these four Majorana fermions into two complex fermions $c_{\uparrow}$ and $c_{\downarrow}$, which implies $n_{\uparrow}+n_{\downarrow}=1$. Therefore, each lattice site's quantum state can be either $\ket{0,1}$or$\ket{1,0}$, so the dimension of the true Hilbert space is $2^L$. 


For the ground state Hamiltonian, rewritten using Majorana fermions, we have:
\begin{eqnarray}
     H_0 =\sum_{\langle i j \rangle \in \gamma}-K u_{ij} (i c_i  c_j)
\end{eqnarray}
where $u_{ij}=i b^{\gamma}_i b_{j}^{\gamma}$ has eigenvalues of $\pm 1$. Noting that $[u_{ij},H_0]=0$, it is a conserved quantity, and since $u_{ij}=(-1)^n$, n is a good quantum number. For periodic boundary conditions, all possible values of $u_{ij}$ have $2^N$ configurations.

On the other hand, the dimension of the fermion Hilbert space before projection is $4^N$, and each $u_{ij}$ has two values. Under periodic boundary conditions, there are $2^N$ different values. However, through $\mathbb{Z}_2$ gauge transformations, these $2^N$ configurations are divided into two equivalence classes: the number of -1s being odd or even, corresponding to anti-periodic boundary conditions and periodic boundary conditions, respectively. Thus, $2^{N-1}$ are not physical, leaving $4^N/2^{N-1}=2^{N+1}$ as the physical Hilbert space.

Given the particle number constraint at each site $n_i=n_{i\uparrow}+n_{i\downarrow}=1$, the total parity of the number of particles is fixed:
$
    (-1)^{\sum_i n_i}=(-1)^N
$
This reduces the Hilbert space from $2^{N+1}$ to $2^{N}$, which becomes the physical Hilbert space.


Under the $\mathbb{Z}_2$ gauge transformation, the bond variable $u_{i j}$ transforms as:
$
    u_{ij} \to -u_{ij}
$.
This means that performing a gauge transformation on site i will flip the sign of both $u_{(i-1) i}$ and $u_{i i+1}$, while leaving the others unchanged. As a result, the parity of the number of -1 values remains unchanged, thereby proving the existence of the two equivalence classes discussed before.

\begin{figure}
    \includegraphics[scale=0.45]{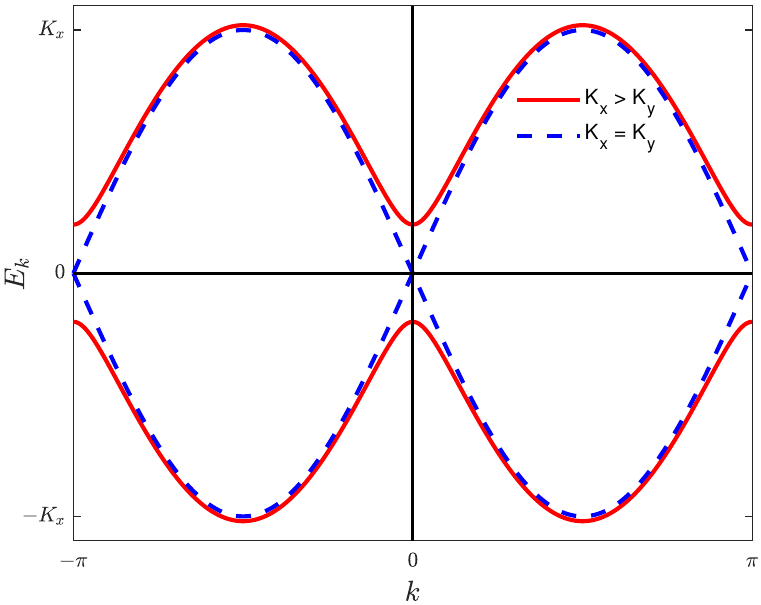}
    \caption{The Dispersion Relation of c Fermion}
    \label{c}
\end{figure}

Applying the mean-field approximation, the ground state Hamiltonian can be written as follows:
\begin{equation}
H_{0} =-K \sum_{\langle i j\rangle \epsilon \gamma}\left[\left\langle u_{i j}\right\rangle\left(i c_{i} c_{j}\right)+\left\langle i c_{i} c_{j}\right\rangle\left(i b_{i}^{\gamma} b_{j}^{\gamma}\right)\right]
\end{equation}
Since $\langle H_0 \rangle <0$, we obtain:
$K\langle u_{ij} \rangle \langle i c_i c_j \rangle >0 $. 
We can fix $\langle i c_i c_j \rangle >0 $. Thus, for $K > 0$, $\langle u_{i j}\rangle=1$ (and $\left\langle u_{i j}\right\rangle=-1$for $K < 0$). Without loss of generality, we assume $K > 0$, and we can easily get the dispersion relations for the $b^x,b^y,b^z,c$ fermions, where the $b$ fermions' are flat, and the $c$ fermions' are shown in Figures \ref{c}.

Considering that the Hamiltonian $H_0$ does not contain $b^z$ terms, there is a ground state degeneracy. Each lattice site has a $b^z$ Majorana fermion. By pairing adjacent sites, we construct$u_{i j}=i b_{i}^z b_{j} ^z$ (with a total of N/2 pairs). From the previous discussion, we know $u_{ij}=\pm 1$. Taking into account periodic and antiperiodic boundary conditions, and the total fermion parity, the ground state degeneracy is given by $2^{\frac{N}{2}+1-1}=2^{\frac{N}{2}}$


Next, we will present the ground state form in the particle number representation:
\begin{equation} \label{groundstate}
    \ket{\begin{array}{lll}
* & *  & * \\
* &  & *
\end{array}}
\end{equation}

Equation \eqref{groundstate} represents the ground state configuration for four adjacent lattice sites. Let these four sites be $k,i,j,l$ in sequence. Without loss of generality, we assume the bonds between them are y-bond, x-bond, and y-bond. The complex fermion formed by two adjacent $b^x$ Majorana fermions is on the x-bond, while the $b^y$ and $b^z$ Majorana fermions are on the y-bond. Thus, the first row of \eqref{groundstate} represents the occupation numbers of the $\gamma$ fermions on the three bonds, and the second row represents the occupation number of the z fermions on the y-bond.

The ground state energies for periodic boundary conditions (PBC) and antiperiodic boundary conditions (APBC) differ, but they converge in the thermodynamic limit. 



\subsection{Perturbative $\Gamma$ terms and the effective model}

 The $\Gamma$ term of Kitaev-Gamma chain for spin-1/2 spins is written as follows:
\begin{eqnarray}
    H'=\sum_{\langle i j\rangle \in \gamma \rm{-bond }}\big[
    \Gamma\big(S_{i}^{\alpha} S_{j}^{\beta}+S_{i}^{\beta} S_{j}^{\alpha}\big)\big]
\end{eqnarray}


In the following we assume that $\Gamma$ is small and will treat the $\Gamma$ interactions as  perturbations: $ H^{\prime}=-\sum_{\langle i j\rangle \in \gamma}\Gamma\left(i b_{i}^{\alpha} b_{j}^{\beta}+i b_{i}^{\beta} b_{j}^{\alpha}\right)\left(i c_{i} c_{j}\right)$.

\subsubsection{Vanishing of first-order Perturbation}
Now, we consider the expectation value of the perturbation $H^{\prime}$ in the ground state,
where $\langle i c_{i} c_{j} \rangle$ is positive and constant. Let's focus on the calculation of
$\langle i b_{i}^{\alpha} b_{j}^{\beta}+i b_{i}^{\beta} b_{j}^{\alpha} \rangle$.

First, consider the term $\gamma=x$ (the term $\gamma=y$ can be handled similarly).
For a fixed x bond, we compute the action of $i b_{i}^{y} b_{j}^{z}$ and $i b_{i}^{z} b_{j}^{y}$ on the ground state separately. We consider the following two cases:



For $K > 0$, take the case where the second line of Equation \eqref{groundstate} is all zero as an example.
\begin{equation}\label{3.8}
    (ib_{i}^{y} b_{j}^{z}+ib_{i}^{z} b_{j}^{y})\ket{\begin{array}{lll}
0 & 0 & 0 \\
0 & &0
\end{array}}=i\lambda_1\ket{\begin{array}{lll}
1 & 0 & 0 \\
0 & & 1
\end{array}}+i\lambda_2\ket{\begin{array}{lll}
0 & 0 & 1 \\
1 & & 0
\end{array}}
\end{equation}
As we all know, the action of $b_{j}^z$ on the j-l bond (the y bond) is $b_{j}^z \ket{0}=\ket{1}$, and the action of $b_{i}^y$ on the k-i bond (the y bond) is $b_{i}^y \ket{0}=-i\ket{1}$. Thus, we obtain $\lambda_1=-i$, and similarly, $\lambda_2=-i$ .

On the other hand, we perform two gauge transformations on the i and j sites, i.e., inserting $b_{i}^x b_{i}^y b_{i}^z c_i=1$ and $b_{j}^x b_{j}^y b_{j}^z c_j=1$ respectively, then we have:
\begin{equation}\label{3.9}
\begin{aligned}
     \ket{\begin{array}{lll}
1 & 0 & 0 \\
0 & & 1
\end{array}}&=- \ket{\begin{array}{lll}
0 & 0 & 1 \\
1 & & 0
\end{array}}
\end{aligned}
\end{equation}
This derivation uses the conclusions $b_i^{x}b_j^{x}=-i$, $c_i c_j \propto -i$, and the properties of majorana fermions, as well as the anticommutation relations of fermions. The detailed process is as follows:


Substituting Equation \eqref{3.9} into Equation \eqref{3.8}, we get:
\begin{equation}
    (ib_{i}^{y} b_{j}^{z}+ib_{i}^{z} b_{j}^{y})\ket{\begin{array}{lll}
0 & 0 & 0 \\
0 & &0
\end{array}}=0
\end{equation}
For other combinations of $b^z$ occupancy, we obtain the same result that the first-order perturbation vanishes.

For $K<0$, we have $u_{ij}=-1,b_i^x b_j^x=i$, and $c_i c_j \propto -i$. In this case, the ground state configuration in the first row consists of all 1s. Then we do the same thing as in the $K > 0$ case and demonstrate that the first-order perturbation does not vanish, lifting the ground state degeneracy and resulting in an energy correction proportional to the absolute value of $\Gamma$. Since perturbation generally lowers the ground state energy, this results in a first-order phase transition at $\Gamma=0$.

Additionally, if the sign of the $\Gamma$ interaction is flipped, the Hamiltonian transforms into a Dzyaloshinskii-Moriya (DM) interaction. This yields the opposite conclusions regarding the perturbative analysis.



\subsubsection{second-order Perturbation and the effective Hamiltonian}
In fact, there is a stronger result: the first-order perturbation of both $\Gamma_x$ and $\Gamma_y$individually is zero. 

We have discussed the second order perturbation of the Gamma interaction, and show that it only contains two $b^z$ terms. Then it is necessary to introduce the following complicated interaction:

\begin{equation}
        H_1=\sum_{\langle i j\rangle ,\langle k l \rangle \in x,j<k} J_{i}\left(S_{i}^{y} S_{j}^{z}+S_{i}^{z} S_{j}^{y}\right) \cdot J_{k} \left(S_{k}^{z} S_{l}^{x}+S_{k}^{x} S_{l}^{z}\right).
\end{equation}
%
We take $H_1$ as 
the new perturbation Hamiltonian. Its expectation value in the ground state subspace is also zero, i.e., the first-order perturbation vanishes. Then, we will consider the second-order perturbation.


In the limit $K_x \gg K_y > 0$, the $c$ fermions open a gap, and their dispersion relation approaches a constant. Therefore, we can assume that the energy required to excite c fermions is the same, making $E_j - E_0=\Delta E$ a constant. Hence, the second-order perturbation simplifies to:
\begin{equation}
    \begin{aligned}
         \sum_{j \ne 0}\frac{\bra{\psi_m} H_1 \ket{\psi_j}\bra{\psi_j} H_1 \ket{\psi_n}}{\Delta E}
         =\frac{\bra{\psi_m} H_1^2 \ket{\psi_n}}{\Delta E}
    \end{aligned}
\end{equation}
When projecting onto the ground state manifold of the Kitaev chain, the effective Hamiltonian only contain 4-fermion interactions, which makes it possible to realize the $q=4$ SYK model.

\section{Implementation of the SYK Model}
\subsection{the SYK Model and its simulation}

The SYK model is characterized by the Hamiltonian:
\begin{equation} \label{H}
    H= (i)^{\frac{q}{2}} \sum_{1 \leq i_{1}<i_{2}<\cdots<i_{q} \leq N} j_{i_{1} i_{2} \cdots i_{q}} \psi_{i_{1}} \psi_{i_{2}} \cdots \psi_{i_{q}} 
\end{equation}
where the coefficients $j_{i_{1} i_{2} \cdots i_{q}}$ are real variables with a random Gaussian distribution. The mean of these coefficients is zero, and their variance is given by:
\begin{equation} \label{gauss}
     \left\langle j_{i_{1} \cdots i_{q}}^{2}\right\rangle=\frac{J^{2}(q-1) !}{N^{q-1}}=\frac{2^{q-1}}{q} \frac{\mathcal{J}^{2}(q-1) !}{N^{q-1}} \quad \text { (no sum) }
\end{equation}

Here, $J$ or $\mathcal{J}$ is a dimensional parameter that ensures the model is more uniformly defined in $q$. We also introduce numerical factors and $N$ factors to simplify the large $N$ limit. For $q\equiv  2$ mod$(4)$, an i factor is introduced to satisfy the Hermiticity of $H$. Consequently, if we limit ourselves to time-reversal symmetric interactions, the $q=4$ model represents the main interaction at low energies.

We aim to implement the $q=4$ SYK model, with the Hamiltonian:
\begin{equation}\label{q=4}
    H=\sum_{iklm}j_{iklm}\psi_{i}\psi_{k}\psi_{l}\psi_{m}
\end{equation}

Now, we focus on the computation of $H_1^2$ mentioned before. We expect that its expansion will contain terms of the form $b_{i}^{z}b_{j}^{z}b_{k}^{z}b_{l}^{z}$. Given the numerous terms in the expansion, we classify them based on the spatial relationship of the interaction terms on different lattice sites. These can be categorized into five types, and it is easy to see that here all types of terms are four $b^z$ terms, where the coefficients $J_{i}J_{k}=J_{i,k},J_{j,k}J_{l,m}=J_{j,k,l,m}$ are real variables satisfying 
   $ \frac{1}{3!}\sum_{k,l,m} \overline{J^{2}_{j,k,l,m}}=\mathcal{J}^{2}$.
And whether they are in the large N limit or not, they can easily satisfy the condition of the average value $\overline{J_{j,k,l,m}}=0 $

In conclusion, $H_1^2$ is actually the sum of all terms of the form $ b_{i}^{z}b_{j}^{z}b_{k}^{z}b_{l}^{z}$, which corresponds to the sum of all four-Majorana fermion interaction terms. As for the coefficients $J_{ik}*J_{j,l}=J_{i,j,k,l}$, this indicates that the $q=4$ SYK model, as given in equation \eqref{q=4}, has an extra spin-spin term.

\subsection{OTOC Calculation}
\subsubsection{the OTOC}
The Out-of-Time-Ordered Correlator (OTOC) is typically defined as:
\begin{equation}\label{CT}
    C_{T}\equiv - \langle[W(t),V(0)]^2 \rangle
\end{equation}

where $\langle \cdots \rangle$ denotes the thermal average (expectation value over the thermal equilibrium state). $W(t)$ and $V(t)$ are operators at time $t$ in the Heisenberg picture. 

We expand \eqref{CT} and write it as a sum of two TOCs and two OTOCs. Among them, the true cause of the increase in the commutator is the behavior of the Out-of-Time-Ordered Correlation (OTOC) function, so we can actually describe chaos through $\langle W(t)^{\dagger} V(0)^{\dagger} W(t) V(0)\rangle$. 
The decrease of OTOC implies an increase in the commutator, hence indicating chaos. It has a close relationship with the scrambling of operators, for example, in the SYK model, the property that local operators do not commute with the Hamiltonian, i.e., $[W,H] \neq 0$, is the reason for the spreading of this operator and the increase of the commutator.\cite{VR2018}

From the general four-point function results, we can write the general OTOC result as: $
    \lambda_{L}=\frac{2 \pi}{\beta}\left(1-\frac{-k^{\prime}(2)}{k_{R}^{\prime}(-1)} \frac{q \alpha_{G}}{\beta \mathcal{J}}+\ldots\right) 
$

Here, $\lambda_{L}$ is the quantum Lyapunov exponent mentioned earlier. When $q=4$, we have:$
    \frac{-k^{\prime}(2)}{k_{R}^{\prime}(-1)} \frac{q \alpha_{G}}{\beta \mathcal{J}} \approx \frac{4.28}{\beta \mathcal{J}} \approx \frac{6.05}{\beta J}$

In this case, the quantum Lyapunov exponent depends only on the temperature T and the coupling coefficient J. Therefore, the OTOC grows exponentially with time.

\subsubsection{Verification in Spin Chain}

In this section, we will numerically verify the exponential growth of the OTOC in the SYK model using spin operators on a spin chain. According to the definition of the OTOC and the construction of the SYK model in a spin chain, we construct spin operators as follows:

For lattice positions $|m-n|\equiv 1 mod(2)$, assuming without loss of generality $m<n$, we consider $S_{m}^{y}(t)S_{m+1}^{z}(0)...S_{n-1}^{z}(0)S_{n}^{y}(0)$, which can be represented using Majorana fermions as:
$b_{m}^{z}(t)b_{n}^{z}(0) (ib_{m}^{x}(t)b_{m+1}^{x}(0))$

If we approximate the effective Hamiltonian of $b_{m}^{x}$ as $h=i b_{m}^{x} b_{m+1}^{x}$, we can analytically calculate $i b_{m}^{x}(t)b_{m+1}^{x}(0)=\mathrm{e}^{-2ti}$.

Combining all the calculations, we can rewrite the OTOC for the SYK model using spin operators as follows:
\begin{equation}
    \begin{aligned}
        \left\langle b_{m}^{z}(t)^{\dagger}b_{n}^{z}(0)^{\dagger} b_{m}^{z}(t)b_{n}^{z}(0) \right\rangle 
        =\langle (S_{m}^{y}(t){S_{m+1}^{z}}...{S_{n-1}^{z}}{S_{n}^{y}})^2     \mathrm{e}^{4ti} \rangle
    \end{aligned}
\end{equation}

For lattice positions $|m-n|\equiv 0 mod(2)$, we simply replace the last spin operator $S_{n}^{y}$ with $S_{n}^{x}$.

Using the Heisenberg representation, we can compute:$
\langle (S_{m}^{y}(t){S_{m+1}^{z}}...{S_{n-1}^{z}}{S_{n}^{y}})^2     \mathrm{e}^{4ti} \rangle$

For numerical computation, we choose L=8, T=0.02, and implement the MATLAB program  to calculate the OTOC. We consider both the spin chain model with and without perturbations.

To visualize the real and imaginary parts of the OTOC separately, we plot them as shown in Figure \ref{yuan}:
\begin{figure}
    \centering
    \subfigure[Without Perturbation]{
    \includegraphics[width=0.45\linewidth]{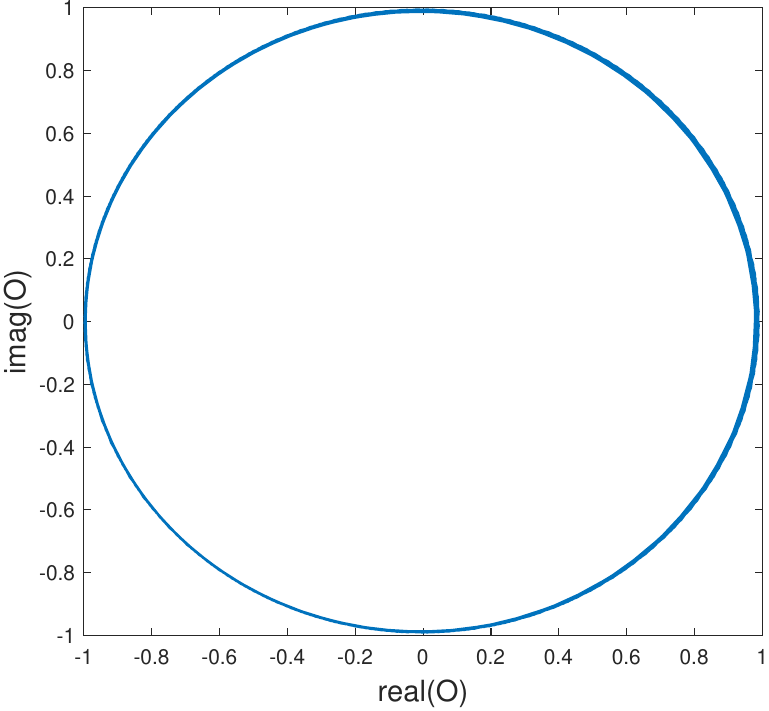}}
    \subfigure[With Perturbation]{
     \includegraphics[width=0.45\linewidth]{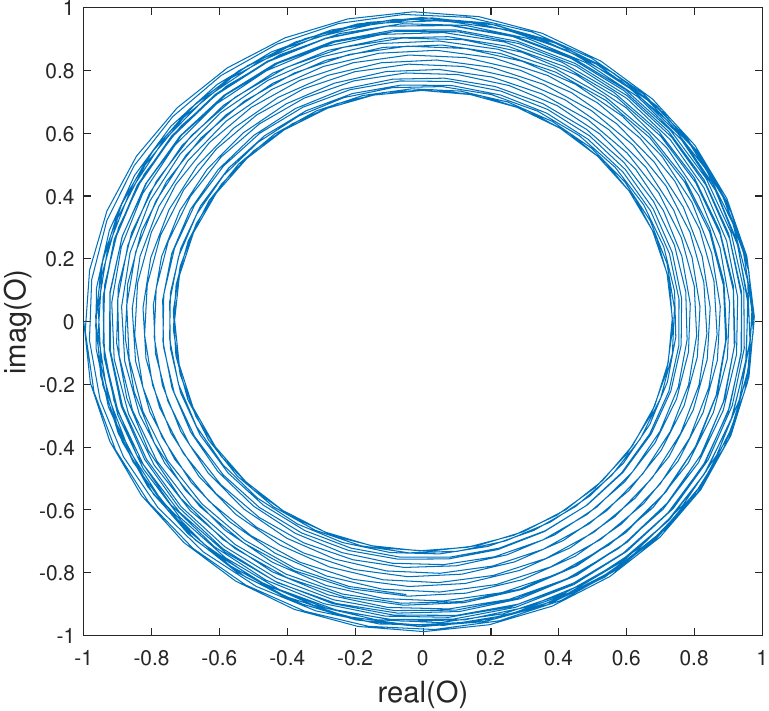}}
    \caption{L=8,T=0.02 Imag(O)-Real(O)}
    \label{yuan}
\end{figure}

To understand why it is not strictly time-independent, it is because in our calculation of the OTOC, we approximated the effective Hamiltonian for $b_{m}^{x}(t)$, while in reality $b^{x}$ and $c$ are correlated.

\begin{figure}[b]
    \centering
    \includegraphics[width=0.45\linewidth]{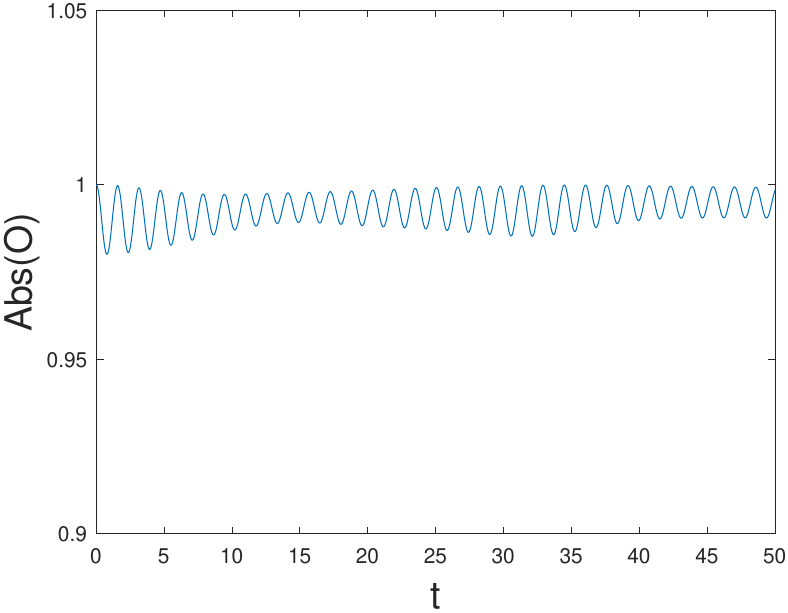}
    \caption{L=8,T=0.02,Without Perturbation}
    \label{wwr}
\end{figure}

When no perturbation is added, the modulus of OTOC is shown in Figure \ref{wwr}. The OTOC magnitude does not evolve significantly with time. This is because without perturbations, the energy of bzbz is zero, hence it does not evolve with time.

 When the perturbation is added, the OTOC calculation results in Figure \ref{logologt} (taking the real part):


In this plot, we observe that in the early time evolution stage, the OTOC appears to grow exponentially, matching the theoretical calculations from the previous section. Taking the logarithm of the OTOC and plotting it over time, as shown in Figure \ref{logologt}:

\begin{figure}[t]
    \centering
    \subfigure[Real(O)-t]{
        \includegraphics[width=0.45\linewidth]{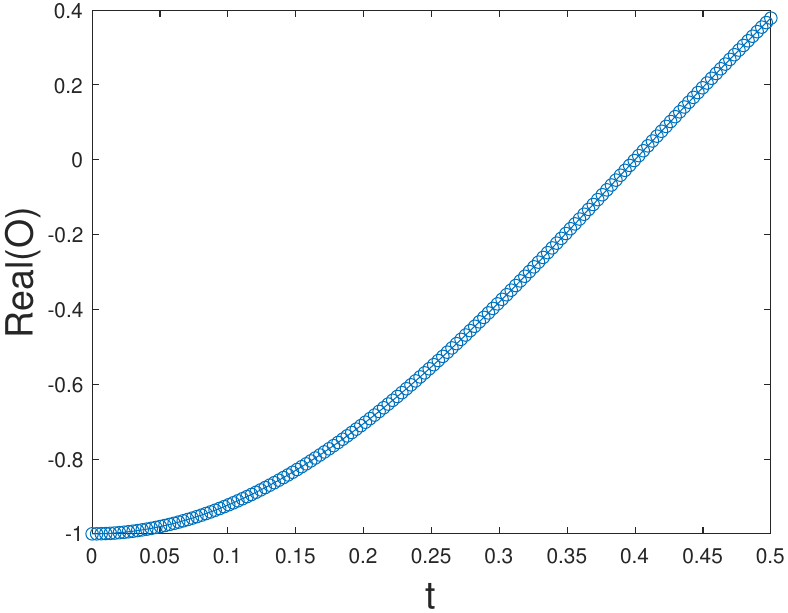}}
    \subfigure[Log(real(O))-Log(t)]{
    \includegraphics[width=0.45\linewidth]{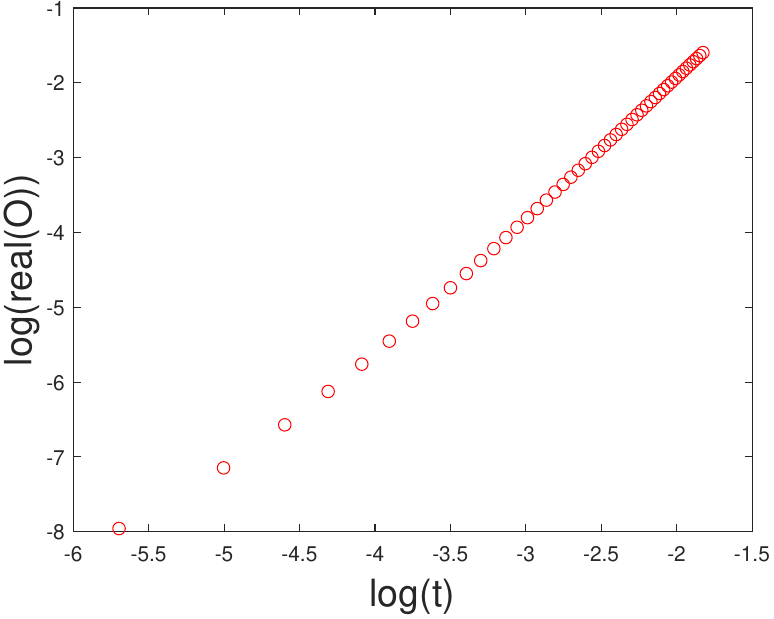}}
    \caption{L=8,T=0.02,With Perturbation}
    \label{logologt}
\end{figure}

Here, we choose data points at intermediate times because at short times, the high-energy contributions are significant, while at long times, the contributions from high-energy parts oscillate and cancel out, leaving mainly low-energy contributions. The computation results seem to show more of a power-law increase rather than an exponential increase. The primary reasons for this deviation could be:\\
\indent (a) Limited Lattice Size: Currently, we can only compute for lattice sizes of 8, 10, or 12, while the SYK model ideally requires a large N limit (i.e., very large lattice size);\\
\indent (b) Finite Temperature: We have not reached the zero-temperature limit in our computations, which might cause deviations from the expected exponential increase;\\
\indent (c) Approximation in Theoretical Calculations: In the theoretical discussion, we made approximations, such as treating the excitation energy of cc fermions as constant and considering $ib_{m}^xb_{m+1}^x$ as a conserved quantity. These approximations might introduce some errors in the numerical results.

Considering these factors, we can still conclude that our results are in agreement with the theoretical predictions.

\section{Conclussions and discussions}

In summary, we propose to realize the SYK model In pure spin systems. Using the Majorana representation of the spins,  we first show that the Kitaev chain contain localized majorana zero modes $b^z$ in the exponentially degenerate ground states. Then we consider the $\Gamma$ term interactions are perturbations. When $K<0$, the first order perturbation is nonzero, hence the energy function has a kink at $\Gamma=0$, indicating a first order phase transition there\cite{}. On the other hand, when $K>0$, the first order perturbation exactly cancels out. Up to the second order perturbation, an effective SYK model with 4-fermion interactions ($q=4$) between the $b^z$ fermions can be realized in the low energy limit. This is an interesting result since it indicates that an effective fermionic model can be simulated in a pure bosonic system.

Furthermore, we illustrate that the OTOC of the majorana fermions can be approximately simulated by the correlators of the spin operators using exact diagonalization. For the pure Kitaev chian where the $b^z$ majorana fermions have precisely zero energy, our simulated OTOC using spin operators is almost independent on time, indicating that it faithfully reflects the OTOC of the majorana fermions. When the $\Gamma$ interactions are turned on, the simulated OTOC is dependent on time, and the short-time behavior at low temperatures is approaching to an exponential function with the increasing length of the spin chain. 

In the pure Kitaev chain, the $b^x$ and $b^y$ majorana fermions are gapped and have almost completely flat dispersion.  The $c$ fermions are enforced to obtain a gap by introducing a dimerized interactions in the amplitude of $K_x$ and $K_y$. When introducing the $\Gamma$ term, the quantity $u_{ij}=ib_{xi}b_{xj}$ in the x-bond and $u_{ij}=ib_{yi}b_{yj}$ are no longer conserved quantities. Hence, strictly speaking the $b^x$ and $b^y$ fermions have non zero contributions to the perturbed ground state. This will introduce extra degrees of freedom and non-SYK type interactions in the effective model. However, when the value of $\Gamma$ is smaller than the gap of the $b^x, b^y$ and $c$ fermions, we can assume that the second order perturbation is a good approximation. 

This work was supported by the Ministry of Science and Technology of China (No. 2022YFA1405300 and No. 2023YFA1406500) the National Natural Science Foundation of China (Grants No. 12374166, No. 12134020).

\appendix

\end{document}